\documentclass[prc,twocolumn,aps,showpacs]{revtex4}
\usepackage{epsfig}
\def\beq{\begin{equation}}
\def\eeq{\end{equation}}
\def\bea{\begin{eqnarray}}
\def\eea{\end{eqnarray}}
\def\<{\langle}
\def\>{\rangle}

\def\gsim{\buildrel > \over {_{\sim}}}
\def\magq{|{\bf q}|}
\usepackage{graphicx}
\usepackage[dvips]{color}
\usepackage{colordvi}
\usepackage{amsfonts}
\usepackage{amssymb}
\usepackage{amsmath}
\usepackage{graphics}
\usepackage{pstricks}
\usepackage{bm}
\usepackage{natbib}

\begin{document}
\title{Final state interactions in the nuclear response at large momentum transfer}
\author{Omar Benhar}
\affiliation{INFN, Sezione di Roma. I-00185 Roma, Italy \\
Dipartimento di Fisica, ``Sapienza" Universit\`a di Roma. I-00185 Roma, Italy}
\date{\today}
\begin{abstract}
The convolution approach, widely employed to describe
final state interactions in the response of 
many-body systems, is derived from the expression of the 
nuclear response in the zeroth-order ladder approximation. Within this framework, 
the folding function, accounting for the effects of interactions between the struck 
particle and the spectator system, can be immediately related to the spectral function of particle states. 
The role of nucleon-nucleon correlations in determining the energy dependence 
is analyzed.
\end{abstract}
\maketitle

\section{Introduction}

In the impulse approximation (IA) regime, the
response of a many-body system to a probe delivering momentum ${\bf q}$
and energy $\omega$, $S({\bf q},\omega)$, can be directly related to the 
spectral function describing
the energy and momentum distribution of its constituents \cite{Benhar01}.

The IA is based on the premise that, as the space resolution
of the probe is $\sim \magq^{-1}$, at large enough $\magq$ (typically 
$\magq \gg 2\pi/d$, $d$ being the average separation between target constituents) the
target is seen by the probe as a collection of individual particles.
In addition, final state interactions (FSI) between the hit constituent
and the spectators are assumed to be negligibly small.

Within the IA scenario, scattering off a many body system reduces to the incoherent sum 
of scattering processes involving the target constituents, the energy and momentum of which 
are distributed according to the spectral function.

The goal of extracting information on the target spectral function from the 
measured cross sections has been pursued in a variety of contexts. The momentum 
distribution of liquid $^4$He, obtained from neutron scattering data, has been 
extensively analyzed to determine the condensate fraction \cite{Mook88}, while a 
number of studies of the nuclear electromagnetic response have been aimed at identifying 
high momentum components, induced by short range nucleon-nucleon (NN) correlations, 
in the target wave function (for a recent review of electron-nucleus scattering in the quasielastic sector,  
see Ref. \cite{Benhar08}).

Comparison between the results of theoretical calculations and 
data has consistently shown that the IA scheme fails to explain the measured 
cross sections at fully quantitative level, thus suggesting that FSI play a significant 
role \cite{Silver88,Meier83,Benhar91,Ciofi92}.
Clearcut evidence of the persistence of FSI effects at large momentum transfer has 
been also provided by theoretical studies of both nuclear matter \cite{Benhar99}
and the Bose hard-sphere system \cite{Negele}.

In view of the fact that FSI may largely obscure the connection between 
the target response and the underlying energy and momentum distribution, a 
quantitative understanding of their effects must be regarded as a prerequisite 
 for the extraction of the relevant dynamical information from the data.

In the widely employed convolution approach, $S({\bf q},\omega)$ is written 
in the form \cite{Silver88,Benhar91}
\beq
\label{convolution}
S({\bf q},\omega) = \int d \omega^\prime \ S_0({\bf q},\omega^\prime) 
 F_{{\bf q}}(\omega-\omega^\prime) \ ,
\eeq
where $S_0({\bf q},\omega)$ denotes the response in the absence of FSI,
the effects of which are described by the folding function $F_{{\bf q}}(\omega)$.

This article is aimed at showing that Eq.(\ref{convolution}), often justified using 
 heuristic arguments \cite{Hohenberg66}, can be obtained in a consistent 
fashion
within a more fundamental approach, based on of many body theory and the 
spectral function formalism.
 \cite{Brown,FetterWalecka}. The proposed interpretation 
 turns out to significantly affect the large $\omega$ behavior of the folding function, that in turn  
determines the tails of the measured inclusive cross section \cite{Benhar91}.

The expression of the nuclear response in terms of spectral functions is derived in 
Section~\ref{foldspec}, while  Section~\ref{eikonal:app} is devoted to the 
discussion of the eikonal approximation, employed to obtain the folding 
function of Eq.~\eqref{convolution}. The elements of the calculation
of $F_{{\bf q}}(\omega)$ are analyzed in Section~\ref{compspec}.
In order to illustrate the role of FSI in determining the electromagnetic nuclear response at large momentum
transfer, in Section~\ref{results} we report the results of theoretical calculations of the inclusive 
electron scattering cross section at $|{\bf q}| \gsim 2 \ {\rm GeV}$. Finally, In 
Section~\ref{conclusions} we summarize our findings and state the conclusions.

\section{Formalism}
\label{formalism}

\subsection{Folding function and particle spectral function}
\label{foldspec}

For simplicity, we will consider a scalar probe interacting with uniform (i.e. translationally invariant) 
isospin symmetric nuclear matter. 
The generalization to the case of finite nuclei and electromagnetic interactions 
does not involve any additional 
conceptual difficulties.  

The response of the system can be written in terms of the  imaginary part
of the particle-hole propagator $\Pi({\bf q},\omega)$ 
according to
\cite{Brown,FetterWalecka,Benhar92}
\beq
\label{def:resp}
S({\bf q},\omega) 
 =  \frac{1}{\pi}\  {\rm Im} \ \Pi({\bf q},\omega) \ ,
\eeq
with
\beq
\Pi({\bf q},\omega) =  \langle 0 \vert \rho^\dagger_{{\bf q}}  \
\frac{1}{H-E_0-\omega-i\epsilon} \ \rho_{{\bf q}} \vert 0 \rangle \ ,
\eeq
and $\epsilon=0^+$. The operator 
\beq
\rho_{{\bf q}}= \sum_{{\bf k}} a^\dagger_{{\bf k}+{\bf q}} a_{{\bf k}}  \ ,
\eeq
$a^\dagger_{{\bf k}}$ and $a_{{\bf k}}$ being  nucleon creation
and annihilation operators, respectively, 
describes the fluctuation of the target density induced by the 
interaction with the probe, while 
the target ground state satisfies the Schr\"odinger equation
\beq
H\vert 0 \rangle = E_0 \vert 0 \rangle \ , 
\eeq
where $H$ denotes the nuclear hamiltonian.

At large momentum transfer, the effects of long range correlations, arising 
from mixing of one particle-one hole states and 
leading to the excitation of collective modes, are expected to become negligible \cite{Benhar09}.
In this kinematical regime, 
the particle-hole propagator can be written in terms of the Green's functions
$G_{h}({\bf k},E)$ and $G_{p}({\bf k},E)$, describing the propagation of 
a nucleon in a hole or particle state with momentum ${\bf k}$ and energy $E$.

The resulting expression of the response, often referred to as zeroth-order ladder approximation  \cite{Benhar92,carlo}, reads
\beq
\label{ladder}
S({\bf q},\omega) = \int d^3k \ dE \ P_h({\bf k},E)P_p({\bf k}+{\bf q},\omega-E) \ ,
\eeq
where $P_h$ and $P_p$ denote the hole and particle spectral functions, respectively, simply related to 
the corresponding Green's functions through \cite{Benhar92,carlo}
\beq
P_{h(p)}({\bf k},E) = - \frac{1}{\pi} \ {\rm Im} \ G_{h(p)}({\bf k},E) \ .
\label{spec:rap}
\eeq

The hole spectral function and the momentum distribution, defined as
\beq
\label{def:momdis}
n({\bf k}) = \int \ dE P_{h}({\bf k},E) \ ,
\eeq 
can be obtained within the framework of non relativistic 
many-body theory. Calculations of $P_h$ in isospin symmetric nuclear matter at equilibrium density 
have been carried out 
using both Correlated Basis Function (CBF) perturbation theory \cite{Benhar89} and 
the self-consistent Green's function approach  \cite{Ramos89}.

Within the IA scheme, in which  FSI are disregarded, the particle spectral function is approximated with that
 of the non interacting Fermi gas. The resulting response
reads
\bea
\nonumber
S_0({\bf q},\omega) & = & \int d^3k \ dE \ P_h({\bf k},E) 
 \theta(|{\bf k}+{\bf q}|-k_F) \\
  & \times & \delta(\omega-E-E_{{\bf k}+{\bf q}}) \ ,
\label{IA}
\eea
where $\theta(x)$ is the Heaviside step function, $k_F$ is the Fermi momentum and $E_{{\bf k}+{\bf q}}$ denotes the kinetic energy of
a nucleon carrying momentum ${\bf k}+{\bf q}$. 

While providing 
an excellent description of the measured nuclear cross sections in the region of the quasielastic peak, the IA scheme leads to
largely underestimate the data at lower energy loss \cite{Benhar08}. In the case of neutron scattering on liquid $^4$He, deviations
from the IA predictions, occurring also at the quasielastic peak, make it difficult the identification of the delta function singularity
associated with the condensate.

In inclusive processes, as long as the set of available final states is complete, FSI do not 
affect the total (i.e. $\omega$-integrated) cross section at fixed ${\bf q}$. At large momentum transfer, their main effect 
is a broadening of the $\delta$-function appearing in 
Eq.(\ref{IA}), owing to the fact that the collisions between the struck particle 
and the spectators couple the one particle-one hole state produced at the primary interaction vertex to more complex final states.
As a result, the state describing the struck particle acquires a finite lifetime 
$\tau \sim 1/\rho \sigma$, where $\sigma$ is the total 
NN scattering cross section  
and $\rho$ is the target density.

The starting point of our derivation is the expression of the response of Eq. \eqref{ladder}, 
providing a link between the particle spectral function and the folding function appearing 
in Eq. \eqref{convolution}. Substituting Eq.~\eqref{IA} into Eq.~\eqref{convolution} and 
comparing the result to the right hand side of Eq.~\eqref{ladder} we find
\bea
\label{rel1}
\nonumber
& & P_p({\bf k}+{\bf q},\omega-E) = \theta(|{\bf k}+{\bf q}|-k_F)  \\
& & \ \ \ \ \ \ \ \ \ \times
\ \int d\omega^\prime F_{{\bf q}}(\omega-\omega^\prime)   
\delta(\omega^\prime-E-E_{{\bf k}+{\bf q}}) .
\eea
At large momentum transfer ($|{\bf q}|~\gg~2~k_F~\sim 500$~MeV in electron-nucleus scattering), the condition that 
the momentum of the struck particle be larger than the Fermi momentum is always 
satisfied and the $\theta$-function can be omitted. Hence, approximating 
${\bf k} + {\bf q} \approx {\bf q}$, which in turn implies 
$E_{{\bf k}+{\bf q}} \approx E_{{\bf q}}$, 
Eqs.(\ref{ladder}) and (\ref{convolution}) become equivalent if 
\beq
\label{rel2}
F_{{\bf q}}(\omega) = P_p({\bf q},\omega + E_{{\bf q}}) \ .
\eeq
Note that the folding function is defined in such a way as to peak at $\omega=0$, 
whereas the spectral function $P_p({\bf q},E)$ is peaked at $E = E_{{\bf q}}$. Furthermore, from 
Eq.(\ref{rel2}) it follows that, since the spectral function is a positive 
quantity \cite{Benhar92}, the folding function $F_{{\bf q}}(\omega)$ is also positive. 


At moderate momentum transfer, the hole and particle spectral functions can be 
consistently obtained using non relativistic many-body theory \cite{Benhar92}. However, in the kinematical
region of large momentum transfer the motion of the struck nucleon in the 
final state can no longer be described using the non relativistic formalism. 

At IA level, the above problem can be easily circumvented replacing the
non relativistic kinetic energy with its relativistic counterpart in Eq.(\ref{IA}).
On the other hand, inclusion of FSI requires further approximations, needed to obtain the 
particle spectral function, or, equivalently, the folding function appearing in Eq.(\ref{convolution}). 

A theoretical approach to calculate the folding function, based on a generalization of Glauber theory of
high energy proton-nucleus scattering \cite{Glauber}, has been 
developed in Ref.~\cite{Benhar91} and extensively applied to the analysis of the measured inclusive 
electron-nucleus cross sections \cite{Benhar93,Benhar94}.

To make a connection between the formalism of Ref.~\cite{Benhar91} and the one based on  
spectral functions, in Section~\ref{eikonal:app} we will outline the derivation of the Green's 
function of particle states within the eikonal approach. 

\subsection{The eikonal approximation}
\label{eikonal:app}

In order to keep contact with the formalism of nuclear many-body theory, 
in this Section we will use the non relativistic formalism. The generalization to 
the case of relativistic particles will be discussed at a later stage. 

Let us first consider a nucleon scattering from the potential $V$.
Its propagation is described by the Green's function
\beq
\label{GF1}
G = (E - {\bf k}^2/2m - V + i \epsilon)^{-1} \ ,
\eeq
that can be obtained solving the integral equation
\beq
\label{GF2}
G = G_0 + G_0 V G \ ,
\eeq
the free space Green's function $G_0$ being given by
\beq
\label{GF0}
G_0 = (E - {\bf k}^2/2m + i \epsilon)^{-1} \ .
\eeq
The eikonal approximation is based on the tenet that, at high energy, the projectile particle travels 
along a straight trajectory with constant speed. Under this assumption, one can write
$E \approx |{\bf p}|^2/2m$, ${\bf p}$ being the incident momentum, and  
the momentum in the intermediate states in the form 
\beq
{\bf k} = {\bf p} + {\bf \Delta} \ \ \ , \ \ \ |{\bf \Delta}| \ll |{\bf p}|  \ .
\eeq
Using the above relations and neglecting the term quadratic in ${\bm \Delta}$, Eq.(\ref{GF0}) can be cast in the form
\beq
\label{eik0}
G_0 = [ v (|{\bf p}| - k_z) + i \epsilon ]^{-1} \ ,
\eeq
where $v = |{\bf p}|/m$ is the nucleon velocity and the $z$-axis has been chosen along the direction of 
${\bf p}$. Fourier transformation to coordinate space yields 
\bea
\nonumber
\label{coord0}
\langle {\bf r}^\prime  | G_0 | {\bf r} \rangle & = & \int \ \frac{d^3 k}{(2 \pi)^3} \  \frac { {\it e^{i {\bf k}\cdot ({\bf r}^\prime-{\bf r})}} } { v (|{\bf p}| - k_z) + i \epsilon   } \\ 
& = & - \frac{i}{v}    \delta( {\bf b}^\prime- {\bf b} )  \theta(z^\prime - z) \ {\it e^{i {|\bf p}| (z^\prime - z)}} \   ,
\eea
where ${\bf b}$ is the projection of ${\bf r}$ on the plane perpendicular to the momentum ${\bf p}$. 

Substituting the above equation in Eq.(\ref{GF2})  and assuming that the interaction $V$ be local, one readily obtains the coordinate-space Green's function
\beq
\label{eik1}
\nonumber
\langle {\bf r}^\prime  | G | {\bf r} \rangle = \langle {\bf r}^\prime  | G_0 | {\bf r} \rangle \ U({\bf b},z^\prime-z) \ ,  
\eeq
with
\beq
\label{eik2}
U({\bf b},z^\prime - z) = 
{\rm exp}  \left[ - \frac{i}{v}  \int_0^{z^\prime-z}  d \zeta \ V({\bf b},z + \zeta) \right]  ,
\eeq
describing the motion of a particle moving along the straight trajectory ${\bf r}(\tau) = {\bf r} + {\bf v} \tau$, 
constrained by the condition ${\bf r}(t)~=~{\bf r}^\prime \equiv({\bf b}, z^\prime)$ at time $t~=~(z^\prime~-~z)/v$.

The above result can be used to obtain
the scattering wave function, $\psi_{{\bf p}}$, from the equation
\beq
\label{def:psi}
\psi_{{\bf p}} = (\openone + GV)\phi_{{\bf p}} \ , 
\eeq
where $\phi_{{\bf p}}$ is an eigenfunction of the free hamiltonian, i.e. a plane-wave of momentum ${{\bf p}}$.
The resulting wave function can be written in the form 
\beq
\psi_{{\bf p}}({\bf b},z) = {\it e}^{ipz} \left[ 1 - \int_{-\infty}^z dz^\prime \Gamma_{{\bf p}}({\bf b}, z^\prime) \right] \ ,
\eeq
with
\beq
\Gamma_{{\bf p}}({\bf b}, z) = V({\bf b},z) \  {\rm exp}  \left[ - \frac{i}{v}  \int_{-\infty}^{z}  d z^\prime \ V({\bf b},z^\prime)\right]  \ .
\label{def:Gamma}
\eeq
Using the above expression of $\psi_{{\bf p}}$ and the 
definition of the scattering amplitude at incident momentum ${\bf p}$ and momentum transfer ${\bf k}={\bf p}-{\bf p}^\prime$ 
(see, e.g., Ref. \cite{sitenko})
\beq
\label{def:ampl}
f_{{\bf p}}({\bf k}) = - \frac{m}{2 \pi} \langle \phi_{{\bf p}^\prime} | V | \psi_{{\bf p}} \rangle \ ,
\eeq
one finds that the quantity defined in Eq.(\ref{def:Gamma}) is trivially related to  $f_{{\bf p}}({\bf k})$ through
Fourier transformation, i.e.
\beq
\label{def2:Gamma}
\Gamma_{{\bf p}}({\bf r}) = - \frac{2 \pi}{m} \int \frac{d^3 k}{(2 \pi)^3} \ {\it e}^{-i {\bf k} \cdot {\bf r}} f_{{\bf p}}({\bf k}) \ .
\eeq 

In the case of a nucleon propagating through nuclear matter in the aftermath of the interaction with an external probe, 
the eikonal approximation must be supplemented with the further assumption that 
the configuration of the spectator system be {\em frozen}, i.e. do not change due to
interactions with the fast projectile particle. 

The eikonal factor including all contributions arising from collisions involving the projectile particle, labelled by the index $1$, and
the $N-1$ target nucleons takes the form
\beq
{\rm exp}  \left[ - \frac{i}{v}   \ \int_0^{z}  d \zeta \sum_{j=2}^N v_{1j}({\bf r}_1 + {\hat {\bf  z}}\zeta - {\bf r}_j) \right]  \ ,
\label{phase:nm}
\eeq
where $v_{ij}$ is the bare NN potential and ${\hat {\bf  z}}$ denotes the unit vector along the $z$-axis.

Expanding the exponential appearing in the right-hand side of Eq.(\ref{phase:nm}), one obtains a series, the terms of which 
are associated with
processes involving an increasing number of interactions between the projectile particle
and the spectator nucleons. The terms corresponding to repeated interactions with the same spectator can be collected and 
summed up to all orders by replacing the bare 
$v_{ij}$ with the coordinate-space scattering amplitude $\Gamma_{{\bf p}}$ of Eq.(\ref{def2:Gamma}).

Average over the nuclear matter ground state leads to the final 
expression of the eikonal factor  [compare to Eq.(\ref{eik2})]
\beq
\label{eik:fact}
U(z) = {\rm exp}  \left[ - \frac{i}{v} \int_0^{z}  d \zeta   \ V(\zeta) \right] \ ,
\eeq
with
\beq
V(\zeta)  = \langle 0 |  \sum_{j=2}^N \Gamma_{{\bf p}}({\bf r}_{1j} + {\hat {\bf z}}\zeta ) | 0 \rangle \ .
\label{nm}
\eeq

Note that the right hand side of the above equation involves an average over the degrees of freedom of both the projectile particle
and the spectators. In general, averaging over the position of the struck particle at $\tau=0$, ${\bf r} _1$,
amounts to a further approximation. However, owing to translation invariance, in uniform nuclear matter this is not the case.

The particle spectral function can be readily computed using Eq.(\ref{spec:rap}) and the 
Green's function obtained within the eikonal approximation 
(for simplicity, from now on the subscript $p$, specifying the 
particle part of both the Green's function and the associated spectral function, will be omitted)
\begin{align}
G({\bf b},z) &= - \frac{i}{v}    \delta({\bf b})  \theta(z)  \\ 
\nonumber
& \times {\rm exp} \left[ i p z - \frac{i}{v} \int_0^{z}  d \zeta   \ V(\zeta)  \right]   \ ,
\end{align}
with $V(\zeta)$ given by Eq.(\ref{nm}).

\section{Calculation of the particle spectral function}
\label{compspec}

\subsection{NN scattering amplitude}

At high incident momentum, ${\bf p}$, the NN scattering amplitude of Eq. \eqref{def:ampl} extracted from the measured cross sections is usually written 
in  terms of three parameters, in the form
\beq
\label{NN:ampl}
f_p({\bf k} )= \frac{p}{4 \pi} \ \sigma_p (\alpha_p + i) \ {\it e}^{-\beta_p {\bf k}^2} \ ,
\eeq
where ${\bf k}$ is the momentum transfer and $\sigma_p$ is the total cross section, while $\alpha_p$ and $\beta_p$ describe the ratio between 
real and imaginary part and the slope, respectively. Note that the above expression fulfills the optical theorem, stating that the forward scattering amplitude
satisfies the relation
\beq
\label{optical}
{\rm Im} \ f_p(0) = \frac{p}{4 \pi} \ \sigma_p \ ,
\eeq
by construction. 

The main effect of FSI, i.e. the broadening of the $\delta$-function appearing in Eq.(\ref{IA}), arises from the 
imaginary part of the scattering amplitude, while the real part produces a shift of the 
response of the order of 10 MeV at most. As we are focusing on a kinematical region in which the typical scale of the energy transfer is several 
hundreds MeV, in the following the effect of the real part of the scattering amplitude will be disregarded, setting $\alpha_p=0$ in  Eq.(\ref{NN:ampl}).  

Medium modifications of the NN scattering cross section are expected to be important, and must be taken into account. In Ref.~\cite{panpiep}, the relation 
between NN scattering in vacuum and in nuclear matter has been analyzed under
the assumption that the nuclear medium mainly affects the 
flux of incoming particles and the phase space available to final state particles, while leaving
the transition probability unchanged. 

The phase space of elastic NN collisions is reduced by Pauli blocking, whereas the modification of the flux is due to the fact that the nucleons involved in the 
scattering process are bound, and therefore off the mass-shell. Within the approach of Ref. \cite{panpiep}, this feature is taken into account by replacing the bare nucleon mass with 
a momentum dependent effective mass. 

\begin{figure}[h!]
\vspace*{.25in}
\includegraphics[width=.90\columnwidth]{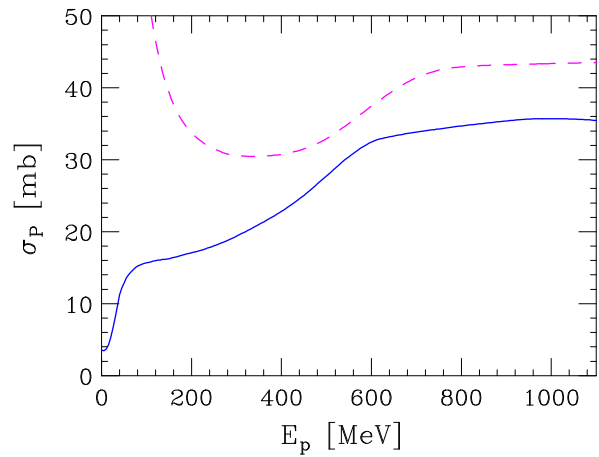}
\caption{{\small (color online) Total proton-neutron cross section as a function of the 
projectile kinetic energy in the Lab frame. The dashed line shows the free space cross section, 
while the solid line has been obtained including medium modifications according to the 
procedure described in Refs. \cite{panpiep,higherord}}.
\label{sigma}}
\end{figure}

The numerical results reported in this article have been obtained using the parametrization of the NN scattering 
amplitude of Eq.(\ref{NN:ampl}), with values of $\beta_p$ and ${\sigma}_p$ taken from the fit of Refs. \cite{oneill1,oneill2}. 
The total cross sections have been corrected for medium effects according to the generalization of the procedure of Ref. \cite{panpiep} 
described in Ref.~\ \cite{higherord}. As shown in Fig. \ref{sigma}, the resulting total cross sections are significantly reduced, with respect to the free space
values. At beam energies $\gsim 800$ MeV the ratio between the proton-neutron cross sections in medium and in vacuum turns out 
to be  $\sim$0.8, and largely energy independent.  

\subsection{Eikonal factor}

Using the expression of the scattering amplitude discussed in the previous Section, 
one can compute the eikonal factor of Eq.(\ref{eik:fact}) with the interaction defined 
by Eq.(\ref{nm}). The ground state expectation value appearing in the right hand side 
can be cast in the form    
\beq
V(\zeta)  = \int d^3r~g(r)~\Gamma_p({\bf r} + {\hat{\bf z}}\zeta) \ ,
\label{nm2}
\eeq
where $g(r)$ is the pair distribution function, yielding the probability of finding
two nucleons separated by a distance $r$ in the nuclear matter ground state.

\begin{figure}[h!]
\vspace*{.25in}
\includegraphics[width=.90\columnwidth]{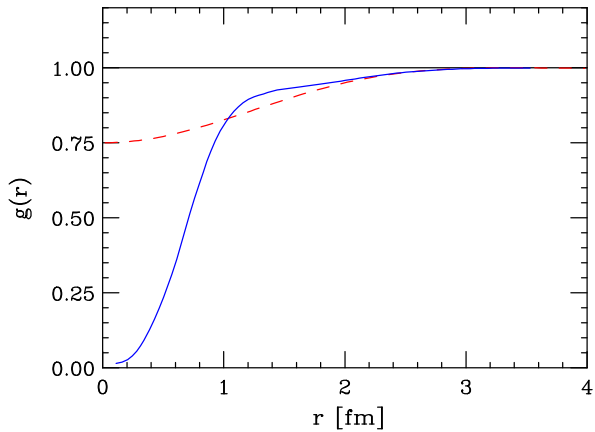}
\caption{{\small (color online) Spin-isospin averaged NN pair correlation function in isospin
symmetric nuclear matter at equilibrium density.
The solid curve shows the result of the many-body calculation of Ref. \cite{compg}, based on a realistic
nuclear hamiltonian, while the dashed curve takes into account statistical correlations only.
The reference line $g(r) \equiv 1$ corresponds to the quasiparticle approximation discussed in the text, in 
which all correlations are neglected. 
\label{g}}}
\end{figure}

The behavior of $g(r)$ is dictated by the strong dynamical NN correlations induced by 
nuclear forces, as well as by the weaker statistical correlations due to Pauli's
exclusion principle. It the absence of all correlations $g(r) \equiv 1$. Realistic
calculations of the nuclear matter pair distribution function have been carried out 
within the Fermi-Hyper-Netted-Chain (FHNC) approach \cite{compg}.
Figure \ref{g} shows a comparison between the results of Ref.~\cite{compg} and
the pair distribution function of the non interacting Fermi gas.

Note that in the simple case of a zero-range coordinate-space interaction, and neglecting 
correlations altogether,  Eq.(\ref{nm2}) yields
\beq
\label{ImV:zerorange}
{\rm Im} \ V = - \frac{1}{2} \rho v \sigma_p \ ,
\eeq
independent of $\zeta$. Note, however, that ${\rm Im} \ V$ depends on density, 
a feature that turns out to be very important in the applications to light nuclei.

\subsection{Spectral function}

Using the results of the previous Sections, the particle spectral function can be obtained from 
Eq.(\ref{spec:rap}), yielding 
\beq
P({\bf p},E) = - \frac{1}{\pi} \ {\rm Im}  \ \frac{i}{v}  
\int_0^\infty dz \ {\it e}^{i (k_z - p) z 
 - \frac{i}{v} \int_0^{z}  d \zeta   \ V(\zeta)}  \ ,
\label{comp:spec}
\eeq
where $p = |{\bf p}|$. Under the assumption that $V$ be purely imaginary, and defining 
\beq
\label{def:W}
W(t) = - \frac{1}{t} \int_0^t \  d\tau  \ {\rm Im}  \ V(v\tau) \ , 
\eeq
where $\tau = \zeta/v$, Eq.(\ref{comp:spec}) can be rewritten 
\beq
P({\bf p},E) = \frac{1}{\pi} \int_0^\infty dt \ \cos [ v(k_z - p) t ] \  {\it e}^{-W(t) t} \ .
\eeq
Note that the right hand side of the above equation depends on the energy $E=E_{\bf k}$ through the 
momentum ${\bf k}$, 
while $W(t)$ depends on $p$ through the momentum-dependence
of the NN scattering amplitude. 

The spectral function takes a very simple form in the case of uncorrelated nucleons
and zero-range scattering amplitude. From Eqs.(\ref{ImV:zerorange}) and \eqref{def:W} it follows that, in this case, 
\beq
\label{def:W0}
W = \frac{1}{2} \rho  v \sigma_p , 
\eeq     
independent of $t$, and 
\beq
\label{P:eikonal}
P({\bf p},E) = \frac{1}{\pi} \frac{ W}{ [v(k_z - p)]^2 + W^2} \ .
\eeq

\subsection{Effects of NN correlations}

The expression of the spectral function of Eq.(\ref{P:eikonal}) deserves some comments.  
Under the assumptions discussed in Section \ref{eikonal:app} $v(k_z - p) \approx E_{{\bf k}} - E_{{\bf p}}$, and 
Eq.(\ref{P:eikonal}) can be rewritten in a form
reminiscent of the definition of the spectral function in terms of the nucleon 
self energy $\Sigma({\bf p},E)$ \cite{Mahaux}
\beq
\label{pke:self}
P({\bf p},E)=\frac{1}{\pi} \frac{ {\rm Im}  \  \Sigma({\bf p},E)  }{ [E-E_{{\bf p}}-{\rm Re} \    \Sigma({\bf p},E) ]^2 + [{\rm Im} \  \Sigma({\bf p},E)]^2}.
\eeq
Comparison between the above equation and Eq.(\ref{P:eikonal}) 
shows that assuming a purely imaginary NN scattering amplitude and neglecting all correlations amounts to 
approximating the self energy with its energy-independent low-density limit, given by the forward NN 
scattering amplitude \cite{Huf1,Huf2}, implying 
\beq
\label{qpa}
{\rm Re} \ \Sigma({\bf p},E) = 0 \ \ \ , \ \ \  
{\rm Im} \ \Sigma({\bf p},E) = \frac{1}{2} \rho v \sigma_p \ .
\eeq
The spectral function obtained within the above approximation
includes the effect of collisions between a nucleon carrying momentum ${\bf p}$
and the nucleons belonging to the Fermi sea. This scheme may be regarded as a {\em quasiparticle approximation}, in 
which only the pole contribution to the Green's function  is taken into account \cite{spect}. 
However, the contribution of NN correlations, resulting in an explicitly energy-dependent self energy, 
is disregarded altogether.  

Within the formalism discussed in this work, the energy dependence associated with correlation 
effects arises from the time dependence of the function $W(t)$, defined by Eq.(\ref{def:W}), 
which is in turn to be ascribed to the time-dependence of the radial distribution function. 

To see this, consider the simple case of a zero-range coordinate-space interaction, i.e.
\beq
{\rm Im}  \ \Gamma_p({\bf r}) = - \frac{1}{2} \rho v \sigma_p \ \delta({\bf r}) \ .
\eeq
Substituting the above expression in Eq.(\ref{nm2}) and using Eq.(\ref{def:W}), one finds 
the result
\beq
\label{def:Wzero}
W(t) =  \frac{1}{2} \rho v \sigma_p \ \frac{1}{t}  \int_0^t d \tau g(v \tau) \ ,
\eeq
which reduces to the time-independent form of Eq.(\ref{def:W0}) for
$g(v \tau) \equiv 1$. Note that within the eikonal approximation time and distance
travelled by the projectile particle are trivially related, as the velocity, $v$, is assumed 
to be constant.

\begin{figure}[h!]
\vspace*{.25in}
\includegraphics[width=.90\columnwidth]{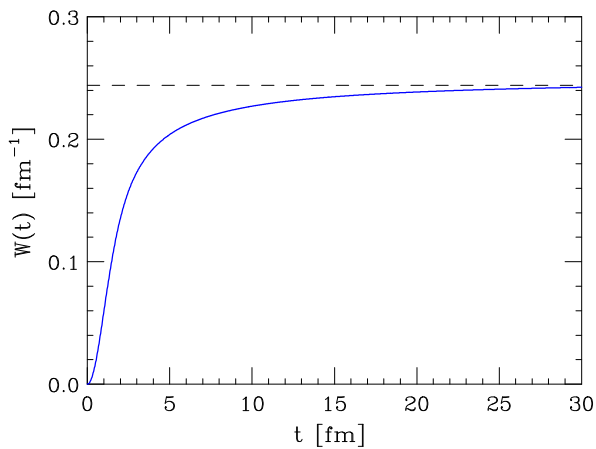}
\caption{{\small (color online) Time-dependence of the function $W(t)$, defined by Eq.(\ref{def:Wzero}), computed using the radial distribution 
function of Fig.~\ref{g} and medium modified NN cross sections obtained from the procedure described in  Refs.~\cite{panpiep,higherord}. 
The nucleon momentum has been set to $p =1 \ {\rm GeV}$. 
\label{W}}}
\end{figure}

The shape of the function W(t), obtained from the above equation using the 
nuclear matter radial distribution function displayed in Fig. \ref{g} and medium modified NN cross sections 
is illustrated in Fig. \ref{W} for the case of a nucleon carrying momentum $p =1 \ {\rm GeV}$.
The deviation from the asymptotic value reflects the fact that, owing to the 
short-range repulsive core of the NN interaction, the struck particle is surrounded by 
a correlation hole, that makes the probability of FSI at short $t$ vanishingly small.  
As a consequence NN correlations mostly affect the high-energy behavior of  
the spectral function.

\section{Results}
\label{results}

In this section we report the results of the application of the formalism discussed in the previous Sections to 
the study of the electromagnetic response of isospin symmetric nuclear matter at momentum transfer 
$|{\bf q}| \gsim 1 \ {\rm GeV}$.

The eikonal approximation derived in the previous Sections to calculate the particle spectral function, yielding in turn the folding function $F_{{\bf q}}(\omega)$,   
can be readily extended to the relativistic regime, relevant to the kinematical region under consideration. It has been shown (see, e.g., Refs. \cite{wallace,jan}) that, 
under the assumptions discussed in Section \ref{eikonal:app}, the relativistic propagator can be linearized and interactions lead to a phase-shift of the plane wave 
describing the motion of the projectile particle. The form of the eikonal phase turns out to be the same as the one given by Eq. \eqref{eik:fact}.

The double differential cross section of the process
\beq
e + A \to e^\prime + X \ ,
\eeq
including the effect of FSI, is obtained convoluting the IA result with the folding function of 
Eq. \eqref{rel2} according to
\beq
\label{xsec}
\frac{d \sigma}{d \Omega d \omega} = \int d \omega^\prime  \left( \frac{d \sigma}{d \Omega d \omega} \right)_{IA} F_{{\bf q}}(\omega - \omega^\prime) \ .
\eeq
The details of the calculation of the IA cross section within the formalism of nuclear many body theory can be found, e.g., in Ref. \cite{Benhar08}. 

\begin{figure}[h!]
\vspace*{.25in}
\includegraphics[width=.90\columnwidth]{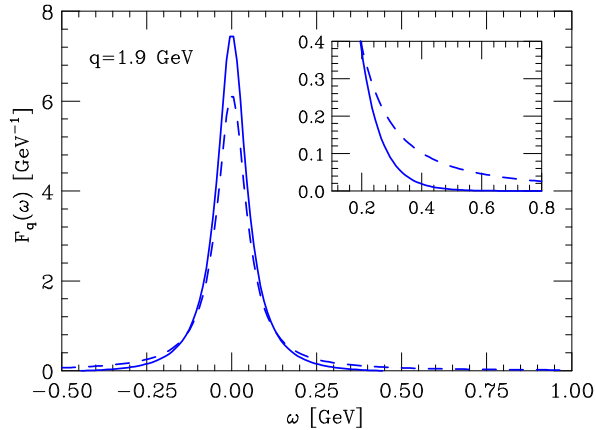}
\caption{{\small (color online) Energy dependence of the folding function defined in Eq. \eqref{rel2}. The solid and dashed line correspond to the full
calculation and to the quasiparticle approximation of Eq. \eqref{qpa}, respectively. The calculations have been carried out for isospin 
symmetric nuclear matter at equilibrium density. The nucleon momentum $|{\bf q}| = 1.9 \ {\rm GeV}$ corresponds to quasi free kinematics at 
incident energy $E_e=3.6 \ {\rm GeV}$ and electron scattering angle $\theta_e = 30 \ {\rm deg}$. 
\label{ff}}}
\end{figure}

As an example, Fig. \ref{ff} shows the behavior of the folding function for incident momentum $\sim 2 \ {\rm GeV}$, computed using 
the radial distribution function of Fig.~\ref{g} and medium modified NN cross sections.
 For simplicity, we have neglected the effect of the finite range of the NN scattering amplitude, which is known to be small \cite{Benhar91}, 
 setting $\beta_p = 0$ in Eq. \eqref{NN:ampl}.
 The solid and dashed lines correspond to the full result and to the quasiparticle approximation [see Eq. \eqref{qpa}], in which NN correlation 
 are neglected altogether. It clearly appears that the inclusion of correlations leads to a reduction of the tails of the folding function, 
 resulting in turn in a reduction of FSI effects in the low-$\omega$ tail of the differential cross section.

\begin{figure}[h!]
\vspace*{.25in}
\includegraphics[width=.90\columnwidth]{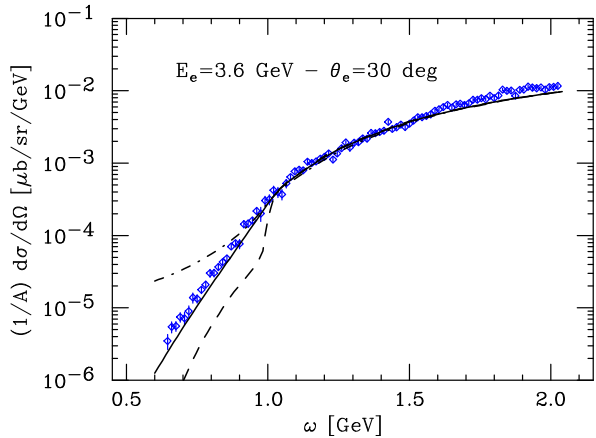}
\caption{{\small (color online) Differential cross section of the scattering process $e~+~A~\to~e^\prime~+~X$ on isospin symmetric nuclear matter, at beam energy 
$E_e = 3.6 \ {\rm GeV}$ and  electron scattering angle $\theta_e = 30 \ {\rm deg}$. The solid and dot-dash lines represent the results of the full calculation and 
those obtained within the quasiparticle approximation discussed in the text, respectively. The cross section obtained within the IA, i.e. neglecting FSI, 
is displayed by the dashed line. The data points show the extrapolated nuclear matter cross section of Ref. \cite{nmdata}.  
\label{dsigma_1}}}
\end{figure}

These features can be observed in Fig. \ref{dsigma_1}, showing the cross section of isospin symmetric nuclear matter 
at beam energy $E_e = 3.6 \ {\rm GeV}$ and electron scattering angle $\theta_e = 30 \ {\rm deg}$. Comparison between the solid and dashed lines, 
corresponding to the full and IA calculations, 
respectively, clearly shows that in the region of low energy loss FSI provide the dominant contribution, which brings theoretical results 
to agree with the extrapolated data of Ref. \cite{nmdata}. The role of NN correlations is illustrated by the dot-dash line, obtained using the folding 
function computed within the quasiparticle approximation (dashed line of Fig. \ref{ff}). It is apparent that neglecting correlations leads to largely 
overestimate FSI effects.
 
\begin{figure}[h!]
\vspace*{.25in}
\includegraphics[width=.90\columnwidth]{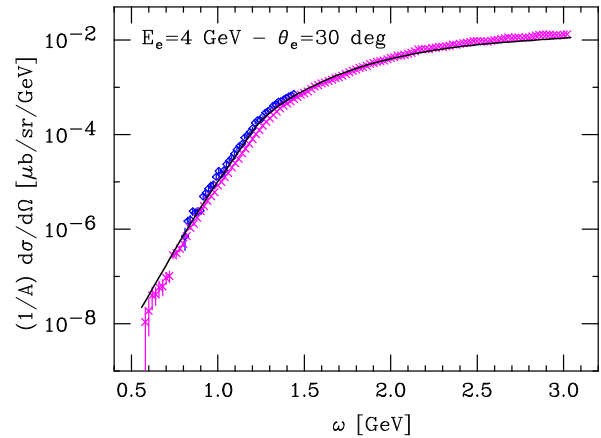}
\caption{{\small (color online) Differential cross section of the scattering process $e~+~A~\to~e^\prime~+~X$ on isospin symmetric nuclear matter, at beam energy $E_e = 4 \ {\rm GeV}$ and  electron 
scattering angle $\theta_e = 30 \ {\rm deg}$. The solid line shows the results of the full calculation, including FSI. 
The diamonds corresponds to the extrapolated nuclear matter cross section of Ref. \cite{nmdata}. For comparison, the crosses also show the cross section of Ref. \cite{newdata}, measured
in the same kinematical setup using a $^{56}Fe$ target.
\label{dsigma_2}}}
\end{figure}

In Fig. \ref{dsigma_2} the differential cross section obtained using the formalism discussed in this article is compared to the extrapolated nuclear matter
data of Ref. \cite{nmdata} and to the $^{56}Fe$ data of Ref. \cite{newdata} at beam energy 
$E_e = 4 \ {\rm GeV}$ and  electron scattering angle $\theta_e = 30 \ {\rm deg}$. The proposed approach appears to provide a quantitative 
description of the measured cross sections over a range exceeding five orders of magnitude.

\section{Conclusions}
\label{conclusions}

We have discussed the description of FSI in the nuclear response, and shown that the widely employed convolution form
of Eq. \eqref{convolution} can be obtained from a fundamental approach based on nuclear many-body theory,  
using the spectral function formalism. 

The folding function of the convolution approach turns out to be directly related to the spectral function describing high momentum nucleons 
occupying particle states, which can be calculated within the eikonal approximation.  The main elements entering this calculation are 
the measured NN scattering cross sections, modified to take into account the effects of the nuclear medium, and the radial distribution function $g(r)$, 
yielding the  probability of finding two nucleons separated by a distance $r$ in the nuclear ground state. Both the nucleon effective mass, driving 
the modifications of the NN cross section, and the radial distribution function are obtained from accurate 
many-body calculations based on a realistic nuclear hamiltonian. 

The interpretation of the folding function as a particle spectral function allows one to unambiguously identify the contribution of 
NN correlations, arising from the energy dependence of the nucleon self-energy.

The main effect of FSI is a redistribution of the strength, producing a slight decrease of the response in the region of the quasi free peak 
and a sharp enhancement of its tails. Inclusion of NN correlations, resulting in the appearance of the energy dependence of the 
nucleon self-energy, leads to a substantial reduction of FSI, to be ascribed to the correlation hole surrounding the struck nucleon.  

The results of numerical calculations of the electron-nucleus scattering cross section at momentum transfer $\gsim 2 \ {\rm GeV}$ show 
that FSI are the dominant reaction mechanism in the region of low energy loss, corresponding to values of the Bjorken scaling variable 
$x \gsim 1.5$. Their effect brings the theoretical results into agreement with the data over a broad range of energy loss.

A pioneering study of the nuclear matter cross section within the convolution approach was carried out in the 1990s \cite{Benhar91}. The 
results of this work suggested that, even after inclusion of NN correlations, using the free space NN cross section to  compute the folding function
leads to sizably overestimate FSI. The authors of Ref.~\cite{Benhar91} argued that the source of this problem could be traced back to 
modifications of the NN cross section arising from the internal structure of the nucleon, and advocated the occurrence color transparency to 
explain the disagreement between theory and data. 

In the present work we have taken into account the modifications of the free space NN cross sections arising from many-body effects, that lead to a 
decrease of FSI. Moreover, unlike that of Ref. \cite{Benhar91}, the folding function resulting from the approach described in this article
is non-negative by definition. The occurrence of oscillations in the tails of the folding function of Ref.~\cite{Benhar91}, 
associated with the appearance of negative values, is incompatible with the interpretation in terms of particle spectral function, and 
as such unphysical. In fact, it must be regarded as numerical noise.

A comprehensive analysis of the large database of inclusive data at large momentum transfer within the present approach will
provide valuable information on the dependence of FSI on both the nuclear mass number, $A$, and the squared four momentum
transfer, $Q^2$. Such a study may help to shed light on the interpretation of the measured ratios of inclusive nuclear cross 
sections at $x >1$  \cite{ratios,ratiosF}. 
 
\acknowledgments
This work was partially supported  by INFN, under grant MB31, and MIUR PRIN, under grant 
``Many-body theory of nuclear systems and implications on the physics of neutron stars''.
The author is deeply indebted to Ingo Sick for a number of illuminating discussions on issues related to the subject of this article, as
well as for a critical reading of the manuscript.

\end{document}